\documentstyle[preprint,prd,aps]{revtex}
\begin{document}
\draft
\title{Thermal Operator Representation of Finite-Temperature 
Amplitudes in the Presence of Chemical Potential} 
\author{M. Inui\footnote{inui@sci.osaka-cu.ac.jp}, H. 
Kohyama\footnote{kohyama@sci.osaka-cu.ac.jp} and A. 
Ni\'{e}gawa\footnote{niegawa@sci.osaka-cu.ac.jp}} 
\address{Graduate School of Science, Osaka City University, 
Sumiyoshi-ku, Osaka 558-8585, JAPAN}
\date{Received today}
\maketitle
\begin{abstract}
In a recent paper [Phys. Rev. D {\bf 72}, 085006 (2005)], Brandt 
{\em et al}. deduced the thermal operator representation for a 
thermal $N$-point amplitude, both in the imaginary-time and 
real-time formalisms. In the case when a chemical potential present, 
however, the representation is not as simple as in the case with 
vanishing chemical potential. We propose a much simpler and 
transparent representation for the case of non-zero chemical 
potential. 
\hspace*{1ex} 
\end{abstract} 
\narrowtext 
\setcounter{equation}{0}
In a recent paper \cite{espi1} (see also \cite{espi2,espi3,espi4}), 
through introducing a {\em thermal operator}, it has been shown that 
any thermal amplitude in the imaginary-time as well as the 
closed-time path formalisms of real-time thermal field theory is 
written in terms of its vacuum-theory counterpart, which is referred 
to as the thermal operator representation. This is proved using the 
mixed space representation $(t, {\bf p})$ with $t$ the time. 
According to the Feynman rules, a thermal $N$-point amplitude is 
computed through following procedure. (1) Draw relevant Feynman 
graphs. (2) For each of them make a product of propagators and 
vertices (in a $(t, {\bf p})$-space) that constitute the graph and 
perform integrations over the internal times, which we write 
$\gamma_N^{(T)}$. Then (3) perform integrations over the loop 
momenta. An important finding in \cite{espi1} is that 
$\gamma_N^{(T)}$ can be factorized as $\gamma_N^{(T)} = {\cal 
O}^{(T)} \gamma_N^{(T = 0)}$, where ${\cal O}^{(T)}$, being 
independent of times, is a thermal operator and $\gamma_N^{(T = 0)}$ 
is the vacuum-theory counterpart of $\gamma_N^{(T)}$. It should be 
noted that, in the case of the imaginary-time formalism, each 
internal-time ($\tau$) integration in $\gamma_N^{(T)}$ is carried out 
over $\int_0^{1/ T} d \tau$, while in $\gamma_N^{(T = 0)}$, each 
integration goes over $\int_{- \infty}^{+ \infty} d \tau$. In the 
case of the closed-time formalism, each internal time ($t$) is 
integrated over $\int_{- \infty}^{+ \infty} d t$, both in 
$\gamma_N^{(T)}$ and $\gamma_N^{(T = 0)}$. Since ${\cal O}^{(T)}$ 
does not depend on the time coordinates, integrations over the 
internal times can be performed within $\gamma_N^{(T = 0)}$. Thus, a 
beautiful factorization of the thermal amplitude into a 
$T$-dependent part (through ${\cal O}^{(T)}$) and a zero-temperature 
part (through $\gamma_N^{(T = 0)}$) is established. 

In \cite{espi1} a generalization of this result to the case where a 
chemical potential $\mu$ exists is attempted. It is found that 
${\cal O}^{(T)}$ involves derivatives with respect to the time 
coordinates, so that the above beautiful factorization property is 
spoiled. 

In this note we present a thermal operator representation for a 
scalar $2 N$-point amplitude at finite temperature and charge 
density ($T \neq 0 \neq \mu$), which is much simpler than the one 
in \cite{espi1} and enjoys the {\em quasi factorizability}. 
\subsubsection*{Imaginary-time formalism} 
In the presence of the chemical potential $\mu$, being conjugate to 
the conserved charge, the complex-scalar propagator in the 
imaginary-time formalism reads 
\begin{eqnarray}
&& \Delta^{(T, \mu)} (\tau, E) \nonumber \\ 
&& \mbox{\hspace*{4ex}} = \frac{1}{2 E} \left[ \theta (\tau) 
\left\{ (1 + n_-) e^{- (E - \mu) \tau} + n_+ e^{(E + \mu) \tau} 
\right\} \right. \nonumber \\ 
&& \mbox{\hspace*{6.5ex}} \left. + \theta (- \tau) \left\{ n_- e^{- 
(E - \mu) \tau} + ( 1 + n_+) e^{(E + \mu) \tau} \right\} \right] 
\, , \nonumber \\ 
&& \mbox{\hspace*{30ex}} (- 1 / T \leq \tau \leq + 1 / T ) \, , 
\label{1.3d} \\ 
&& \Delta^{(T = 0 = \mu)} (\tau, E) \nonumber \\ 
&& \mbox{\hspace*{4ex}} = \frac{1}{2 E} \left[ \theta 
(\tau) e^{- E_- \tau} + \theta (- \tau) e^{E_+ \tau} \right]_{E_\pm 
= E} \nonumber \\ 
&& \mbox{\hspace*{30ex}} (- \infty < \tau < + \infty) \, , 
\label{Tmu0} 
\end{eqnarray}
where $\tau$ is the Euclidean time, $T$ is the temperature, and 
\[ 
n_\pm = n (E \pm \mu) \, , 
\] 
with $n (x) = 1 / (e^{x / T} - 1)$. For later convenience, in Eq. 
(\ref{Tmu0}), we have introduced $E_\pm$. Now, we introduce the 
operators $R (E)$, ${\cal N}$, and $S (\mu)$: $R (E)$ is a \lq\lq 
twisted'' reflection operator that changes $E_\pm \to - E_\mp$ 
(namely, it gives a term with $E_\pm \to - E_\mp$), ${\cal N}$ is an 
operator that acts as ${\cal N} e^{\pm E_\pm \tau} = n (E_\pm) 
e^{\pm E_\pm \tau}$, and $S (\mu)$ is a translation operator that 
changes $E_\pm \to E \pm \mu$. In contrast to $S (E)$ in 
\cite{espi1}, $R (E)$ here does not act on the factor $1 / (2 E)$ 
in Eq. (\ref{Tmu0}). Then, one can easily find the following 
representation (cf. Eqs. (\ref{1.3d}) and (\ref{Tmu0})); 
\begin{eqnarray}
{\Delta}^{(T, \mu)} & = & S (\mu) \left[ 1 + {\cal N} ( 1 + 
R (E)) \right] \Delta^{(T = 0 = \mu)} \, , \nonumber \\ 
& \equiv & {\cal O}^{(T, \mu)} (E, \mu) \Delta^{(T = 0 = \mu)} \, . 
\label{def}
\end{eqnarray}

Let $A^{(2N)}$ be a complex-scalar $2 N$-point amplitude. Following 
the same procedure as in \cite{espi1}, we arrive at  
\begin{eqnarray}
A^{(2 N)} &= & \int \prod_{i = 1}^I \frac{d^{\, 3} k_i}{(2 \pi)^3} 
\prod_{v = 1}^V (2 \pi)^3 \delta_v ({\bf k}, {\bf p}) \gamma_{2 
N}^{(T, \mu)} \nonumber \\ 
& = & \int \prod_{i = 1}^I \frac{d^{\, 3} k_i}{(2 \pi)^3} 
\prod_{v = 1}^V (2 \pi)^3 \delta_v ({\bf k}, {\bf p}) {\cal O}^{(T, 
\mu)} \gamma_{2 N}^{(T = 0 = \mu)} \, , \nonumber \\ 
&& 
\label{espfin} 
\end{eqnarray}
where, with obvious notation, 
\begin{eqnarray*}
{\cal O}^{(T, \mu)} & = & \prod_{i = 1}^I {\cal O}^{(T, \mu)}_i 
(E_i, \mu) \\ 
& \equiv & \prod_{i = 1}^I S_i (\mu) \left[ 1 + {\cal N}_i (1 
+ R (E_i)) \right] \, . 
\end{eqnarray*}
Here $I$ $(V)$ is the total number of internal propagators 
(vertices) in the graph under consideration. The internal and 
external three momenta are denoted generically by $({\bf k}, {\bf 
p})$, respectively, and $\delta_v ({\bf k}, {\bf p})$ enforces the 
three-momentum conservation at the vertex $v$. As mentioned at the 
beginning, the range of the internal-time integration in $\gamma_{2 
N}^{(T, \mu)}$ is $(0, 1 / T)$, while the range of the integration 
in $\gamma_{2 N}^{(T = 0 = \mu)}$ is $(- \infty, + \infty)$. 

Each term $F \in \gamma_{2 N}^{(T = 0 = \mu)}$ includes $E_{i +}$ or 
$E_{i -}$; $F (..., E_{i +}, ... )$ or $F (..., E_{i -}, ... )$. 
>From the above construction, actions of ${\cal O}_i^{(T, \mu)}$ on 
$F$ are unambiguously defined: 
\begin{eqnarray}
(1 + {\cal N}_i) F (..., E_{i \pm}, ... ) & = & (1 + n (E_{i \pm})) 
F (..., E_{i \pm}, ... ) \, , \nonumber \\ 
{\cal N}_i R (E_i) F (..., E_{i \pm}, ... ) & = & n (E_{i \mp}) F 
(..., - E_{i \mp}, ... ) \, . 
\label{kisoku}
\end{eqnarray}

The heart of the proof of Eq. (\ref{espfin}) resides in the proposition: 
\begin{description}
\item{}
The function $G$ defined by 
\begin{eqnarray*}
G & \equiv & \int_{- \infty}^{+ \infty} d \tau \left( \theta (- 
\tau) + \theta (\tau - 1 / T) \right) \nonumber \\ 
&& \times \prod_{i = 1}^N \Delta^{(T = 0 = \mu)} (\tau - \tau_i, 
E_i) \nonumber \\ 
&& \times \prod_{j = N + 1}^{2 N} \Delta^{(T = 0 = \mu)} (\tau_j - 
\tau, E_j) \, , 
\end{eqnarray*}
which appears in complex-scalar  $\lambda (\phi^\dagger \phi)^N$ 
theory, is annihilated by ${\cal O}^{(T, \mu)}$, ${\cal O}^{(T, 
\mu)} G = 0$. 
\end{description}
Proof: Substituting the expression (\ref{Tmu0}), we obtain 
\begin{eqnarray*}
G & = & \prod_{i = 1}^{2 N} \frac{1}{2 E_i} \left( 
\frac{\prod_{i=1}^N e^{- E_{i +} \tau_i} \prod_{j= {N + 1}}^{2 N} 
e^{- E_{j -} \tau_j}}{\sum_{i = 1}^N E_{i +} + \sum_{j = N + 
1}^{2 N} E_{j -} } \right. \nonumber \\ 
&& \left. + \frac{\prod_{i=1}^N e^{- E_{i -} / T} e^{E_{i -} 
\tau_i} \prod_{j= {N + 1}}^{2 N} e^{- E_{j +} / T} e^{E_{j +} 
\tau_j}}{\sum_{i = 1}^N E_{i -} + \sum_{j = N + 1}^{2 N} E_{j +} } 
\right) \, . 
\end{eqnarray*}
Straightforward manipulation using the rules (\ref{kisoku}) and the 
identity $n (E_\pm) e^{E_\pm / T} = 1 + n (E_\pm)$ shows that 
\[
{\cal O}^{(T, \mu)} G = \prod_{i = 1}^N {\cal O}^{(T \, \mu)}_i 
(E_i, \mu) G = 0 \, . \mbox{\hspace*{10ex} (Q.E.D.)} 
\]

Using this proposition, the \lq\lq general proof'' in \cite{espi1} goes 
as it is in the present case, and Eq. (\ref{espfin}) is proved. 

As seen above in conjunction with Eq. (\ref{kisoku}), actions of 
${\cal O}^{(T, \mu)}$ on the \lq\lq $E_+$ sectors'' and \lq\lq $E_-$ 
sectors'' of $\gamma^{(T = 0 = \mu)}_{2 N}$ are asymmetric and, in 
contrast to the case of vanishing chemical potential, Eq. 
(\ref{espfin}) does not achieve the \lq\lq complete factorization''. 
Nevertheless, as seen above, computational procedure on the basis of 
Eq. (\ref{espfin}) of a thermal amplitude from its zero-temperature 
counterpart is unambiguously defined. 
\subsubsection*{Closed-time-path formalism} 
The {\em quasi factorizability} established above for the 
imaginary-time formalism holds also in the closed-time-path 
formalism of real-time thermal field theory. The propagator enjoys 
$(2 \times 2)$-matrix structure, whose $(i, j)$-element reads 
\begin{eqnarray*}
\Delta_{11 (22)}^{(T, \mu)} (t, E) & = & \frac{1}{2 E} \left[ 
\theta (\pm t) e^{- i (E - \mu \mp i \epsilon) t} \right. \nonumber 
\\ 
&& + \theta (\mp t) e^{i (E + \mu \mp i \epsilon) t} + n_- e^{- i 
(E - \mu) t} \nonumber \\ 
&& \left. + n_+ e^{i (E + \mu) t} \right] 
\, , \nonumber \\ 
\Delta_{1 2}^{(T, \mu)} (t, E) & = & \frac{1}{2 E} \left[ 
n_- e^{- i (E - \mu) t} + (1 + n_+) e^{i (E + \mu) t} \right] \, , 
\nonumber \\ 
\Delta_{2 1}^{(T, \mu)} (t, E) & = & \frac{1}{2 E} \left[ 
( 1 + n_-) e^{- i (E - \mu) t} + n_+ e^{i (E + \mu) t} \right] \, . 
\nonumber 
\end{eqnarray*}
The suffices \lq 1' and \lq 2' are called thermal indices. The 
vacuum-theory counterparts are 
\begin{eqnarray*}
\Delta_{11 (22)}^{(T = 0 = \mu)} (t, E) & = & \frac{1}{2 E} \left[ 
\theta (\pm t) e^{- i (E_- \mp i \epsilon) t} \right. \nonumber \\ 
&& \left. + \theta (\mp t) e^{i (E_+ \mp i \epsilon) t} 
\right]_{E_\pm = E} \, , \nonumber \\ 
\Delta_{1 2 (2 1)}^{(T = 0 = \mu)} (t, E) & = & \frac{1}{2 E} 
e^{\pm i E_\pm t} \, \rule[-2.8mm]{.14mm}{7.5mm} 
\raisebox{-2.0mm}{\scriptsize{$\; E_\pm = E$}} \, . 
\end{eqnarray*}
We modify the definition of ${\cal N}$ as 
\begin{eqnarray*}
{\cal N} e^{- i (E_- \pm i \epsilon) t} &=& n (E_-)  e^{- i (E_- \pm 
i \epsilon) t} \, , \nonumber \\ 
{\cal N} e^{i (E_+ \pm i \epsilon) t} &=& n (E_+) e^{i (E_+ \pm i 
\epsilon) t} \, . 
\end{eqnarray*}
Then we have 
\begin{eqnarray*}
\Delta_{i j}^{(T, \mu)} (t, E) &=& {\cal O}^{(T, \mu)} (E, \mu) 
\Delta_{i j}^{(T = 0 = \mu)} (t, E) \, . 
\end{eqnarray*}
Here ${\cal O}^{(T, \mu)}$ is as in Eq. (\ref{def}) with the above 
modification for ${\cal N}$. This quasi factorizability of the 
propagator straightforwardly leads, in much the same manner as in 
\cite{espi1}, to the thermal operator representation for the 
contribution to the amplitude from any graph. 
\subsubsection*{Comment on the general real-time contour} 
Finally, we make a comment on the formalism constructed employing a 
general real-time contour, $- \infty \to + \infty \to + \infty - i 
\sigma / T \to - \infty - i \sigma / T \to - \infty - i / T$ ($0 
\leq \sigma \leq 1$), in a complex time plane. Let ${\cal A}_{i_1 
i_2 ... i_{2 N}}^{(\sigma)} (p_1, ... , p_N; p_{N + 1}, ... , 
p_{2N})$ be a $2 N$-point amplitude in a energy-momentum space 
computed in this formalism, where the suffices $i$'s, each of which 
is 1 or 2, are the thermal indices. The closed-time-path counterpart 
is ${\cal A}_{i_1 i_2 ... i_{2 N}}^{(\sigma = 0)} (p_1, ... , 
p_N; p_{N + 1}, ... , p_{2 N})$. Here all four-momenta are defined 
to be incoming. It can easily be shown that 
\begin{eqnarray}
&& {\cal A}_{i_1 i_2 ... i_{2 N}}^{(\sigma)} (p_1, ... , p_N; p_{N 
+ 1}, ... , p_{2 N}) \nonumber \\ 
&& \mbox{\hspace*{4ex}} = e^{\sigma \sum_{j} p_j^0 / T} 
{\cal A}_{i_1 i_2 ... i_{2N}}^{(\sigma = 0)} (p_1, ... , p_N; p_{N 
+ 1}, ... , p_{2 N}) \, , \nonumber \\ 
&& 
\label{dasoku}
\end{eqnarray}
where the summations are taken over all \lq\lq type 2'' external 
vertices and $p_j^0$ is the $0$th component of $p_j$. Then, ${\cal 
A}^{(\sigma)}$ can always be directly obtained from ${\cal 
A}^{(\sigma = 0)}$ and separate analysis of ${\cal A}^{(\sigma \neq 
0)}$ is {\em practically} not necessary. Incidentally, from Eq. 
(\ref{dasoku}), we see that ${\cal A}_{11 ... 1}^{(\sigma)} = {\cal 
A}_{11 ... 1}^{(\sigma = 0)}$, as it should be. Using the energy 
conservation, $\sum_{i = 1}^{ 2N} p_i^0 = 0$, we have ${\cal A}_{22 
... 2}^{(\sigma)} = {\cal A}_{22 ... 2}^{(\sigma = 0)}$, as also it 
should be. 

A. N. is supported in part by the Grant-in-Aid for Scientific 
Research [(C)(2) No. 17540271] from the Ministry of Education, 
Culture, Sports, Science and Technology, Japan, No.(C)(2)-17540271. 
 
\end{document}